\documentclass[apj,twocolumn]{openjournal}
\usepackage{amsmath}
\usepackage{booktabs}

\usepackage[dvipsnames]{xcolor} 
\usepackage[breaklinks,colorlinks,citecolor=blue,urlcolor=blue,linkcolor=blue]{hyperref}

\usepackage{orcidlink}
\usepackage{subfigure}

\renewcommand{\arraystretch}{1.2}  
\usepackage{listings}
\usepackage{color}

\usepackage{soul}
\usepackage{amsmath}
\usepackage{xspace}
\usepackage{graphicx}
\usepackage{enumitem}
\usepackage{amssymb}
\usepackage{xifthen}
\usepackage{hyperref}
\usepackage[normalem]{ulem}
\usepackage{multirow}

\usepackage[hang,flushmargin]{footmisc} 

\usepackage{styles/todo_notes}
\usepackage{styles/abbreviations}

\makeatletter
\newcommand{\unit}[1]{%
    \,\mathrm{#1}\checknextarg}
\newcommand{\checknextarg}{\@ifnextchar\bgroup{\gobblenextarg}{}}
\newcommand{\gobblenextarg}[1]{\,\mathrm{#1}\@ifnextchar\bgroup{\gobblenextarg}{}}
\makeatother

\newif\ifstartedinmathmode
\newcommand{\msun}{%
  \relax\ifmmode\startedinmathmodetrue\else\startedinmathmodefalse\fi
  {\ifstartedinmathmode\unit{M_{\odot}}\else$\unit{M_{\odot}}$\fi}\xspace%
}

\newif\ifstartedinmathmode
\newcommand{\rsun}{%
  \relax\ifmmode\startedinmathmodetrue\else\startedinmathmodefalse\fi
  {\ifstartedinmathmode\unit{R_{\odot}}\else$\unit{R_{\odot}}$\fi}\xspace%
}

\makeatletter
\renewcommand\@makecaption[2]{%
  \par
  \vskip\abovecaptionskip
  \begingroup
    \footnotesize\rmfamily
    \begingroup
      \samepage
      \flushing
      \let\footnote\@footnotemark@gobble
      \ifnum\pdfstrcmp{\@captype}{table}=0
        \@make@capt@title{\textsc{Table \thetable}}{#2}%
      \else
        \ifnum\pdfstrcmp{\@captype}{figure}=0
          \@make@capt@title{\textsc{Figure \thefigure}}{#2}%
        \else
          \@make@capt@title{#1}{#2}%
        \fi
      \fi\par
    \endgroup
  \endgroup
  \vskip\belowcaptionskip
}
\makeatother

\defcitealias{Kiel+2009:2009MNRAS.395.2326K}{KH09}

\newcommand{\referee}[1]{#1}

\begin{document}

\author{Tom Wagg\,\orcidlink{0000-0001-6147-5761}$^{1}$}
\author{David~D.~Hendriks\,\orcidlink{0000-0003-0872-7098}$^{2}$}
\author{Mathieu Renzo\,\orcidlink{0000-0002-6718-9472}$^{3}$}
\author{Katelyn Breivik\,\orcidlink{0000-0001-5228-6598}$^{4}$}

\affiliation{$^1$Department of Astronomy, University of Washington, Seattle, WA, 98195, USA}
\affiliation{$^2$Department of Physics, University of Surrey, Guildford GU2 7XH, Surrey, UK}
\affiliation{$^3$University of Arizona, Department of Astronomy \& Steward Observatory, 933 N. Cherry Ave., Tucson, AZ 85721, USA}
\affiliation{$^4$McWilliams Center for Cosmology and Astrophysics, Department of Physics, Carnegie Mellon University, Pittsburgh, PA 15213, USA}

\email{Corresponding author: tomjwagg@gmail.com}
\title{Stellar ejection velocities from the binary supernova scenario:\\A comparison across population synthesis codes}

\begin{abstract}
    The vast majority of binary systems are disrupted at the moment of the first supernova, resulting in an unbound compact object and companion star. These ejected companion stars contribute to the observed population of runaway stars. Therefore, an understanding of their ejection velocities is essential to interpreting observations, particularly in the \textit{Gaia} era of high-precision astronomy.
    We present a comparison of the predicted ejection velocities of disrupted binary companions in three different population synthesis codes: \cosmic, \compas and \binc, which use two independent algorithms for the treatment of natal kicks. We confirm that, despite the codes producing different pre-supernova evolution from the same initial conditions, they each find the ejection velocities of secondary stars from disrupted binaries are narrowly distributed about their pre-supernova orbital velocity. We additionally include a correction to the derivation included in \citet{Kiel+2009:2009MNRAS.395.2326K} that brings it into agreement with methods from other works for determining post-supernova binary orbital parameters. During this comparison, we identified and resolved bugs in the kick prescriptions of \textit{all three} codes we considered, highlighting how open-science practices and code comparisons are essential for addressing implementation issues.
\end{abstract}

\maketitle
\section{Introduction}

Observations of pulsars in the Milky Way indicate that they have a much higher velocity dispersion than their massive star progenitors, and hence that neutron stars acquire high velocities upon formation \citep[e.g.,][]{Gunn+1970:1970ApJ...160..979G, Hobbs+2005:2005MNRAS.360..974H, Igoshev+2020}. For compact objects formed via isolated binary evolution, these velocities are expected to be imparted at the moment of core-collapse. Even for a symmetric explosion, the instantaneous mass loss from a supernova (SN) changes the post-SN orbital parameters \citep{Blaauw+1961, Boersma+1961}. The symmetric mass loss alone is typically insufficient to disrupt the orbit, since the mass that can be ejected is often limited by loss during previous mass transfer phases. However, in some cases (${\sim}16\%$) symmetric mass loss from supernova ejecta alone can unbind the binary \citep{Renzo+2019:2019A&A...624A..66R}. This is typically only possible for wide binaries that have massive SN progenitors, which eject a significant fraction of their mass. Moreover, asymmetry in the explosion can impart a significant natal kick on the newly formed compact object \citep[e.g.,][]{Shklovskii+1970:1970SvA....13..562S,Lyne+1994:1994Natur.369..127L,Janka+2013:2013MNRAS.434.1355J,Janka+2017:2017ApJ...837...84J, Burrows+2024:2024ApJ...963...63B}, which in the majority of cases is strong enough to disrupt a binary \citep[e.g.,][]{DeDonder+1997:1997A&A...318..812D,Eldridge+2011:2011MNRAS.414.3501E,Renzo+2019:2019A&A...624A..66R}. Similar asymmetry in neutrino emission can result in additional natal kicks, which can dominate in the case of full fallback \citep[e.g.,][]{Janka+1994:1994A&A...290..496J,Burrows+1996:1996PhRvL..76..352B,Scheck+2006:2006A&A...457..963S,Wongwathanarat+2013,Coleman+2022:2022MNRAS.517.3938C,Vigna-Gomez+2024:2024PhRvL.132s1403V}.

The effect of these SN kicks on the post-SN orbital parameters of a binary system has been considered in detail by many previous works \citep[e.g.,][]{Flannery+1975:1975A&A....39...61F,Sutantyo+1978:1978Ap&SS..54..479S,Hills+1983:1983ApJ...267..322H,Dewey+1987:1987ApJ...321..780D,Wijers+1992:1992A&A...261..145W,Brandt+1995:1995MNRAS.274..461B,Kalogera+1996:1996ApJ...471..352K,Tauris+1998:1998AA...330.1047T,Hurley+2002,pfahl_comprehensive_2002,Belczynski+2008:2008ApJS..174..223B,Kiel+2009:2009MNRAS.395.2326K, Pijloo+2012:2012MNRAS.424.2914P}. These investigations account for a variety of effects, with the most extensive models addressing the impact of the natal kick on the compact object, including the effect of instantaneous symmetric mass loss, handling eccentric pre-SN systems, the probability of disruption of the system, the SN blast wave impulse on the secondary star \citep[e.g.,][]{Liu+2015:2015A&A...584A..11L, Hirai+2018:2018ApJ...864..119H, Ogata+2021:2021MNRAS.505.2485O,Wong+2024:2024ApJ...973...65W}, and collisions between the compact object and companion star \citep[e.g.,][]{Davies+1992:1992ApJ...401..246D,Hirai+2022:2022MNRAS.517.4544H, Kremer+2022:2022ApJ...933..203K}. For a detailed derivation, which explains how each of these effects impacts a binary in clear and compact vector notation, we refer the reader to Appendix B of \citet{pfahl_comprehensive_2002}. 

SN kicks have a wide-ranging impact on massive stellar populations, for both disrupted and bound binary products. The post-SN eccentricity, semi-major axis and centre-of-mass velocity of bound systems, as well as the rate of unbinding and ejection velocities of stars and compact objects, are sensitive to modelling choices in kick prescriptions. The demographics and rates of bound binaries containing at least one compact object have been shown to be strongly affected by the magnitude and distribution of SN kicks. In particular, kicks affect the prevalence of galactic X-ray binaries \citep[e.g.,][]{Tauris+1999:1999MNRAS.310.1165T, Pfahl+2002:2002ApJ...574..364P, Wong+2012:2012ApJ...747..111W,Vigna-Gomez+2024:2024PhRvL.132s1403V}, dormant black hole binaries \citep[e.g.,][]{Breivik+2017:2017ApJ...850L..13B}, double neutron stars \citep[e.g.,][]{Brandt+1995:1995MNRAS.274..461B,Wex+2000:2000ApJ...528..401W,Podsiadlowski+2004:2004ApJ...612.1044P,Bray+2016:2016MNRAS.461.3747B, Vigna-Gomez+2018:2018MNRAS.481.4009V}, and gravitational-wave sources \citep[e.g.,][]{Fryer+1998:1998ApJ...496..333F,Belczynski+1999:1999A&A...346...91B,Dominik+2013:2013ApJ...779...72D,Broekgaarden+2022, Wagg+2022}. Moreover, the observed population of runaway stars has grown significantly in the era of \textit{Gaia} \citep[e.g.,][]{Carretero-Castrillo+2023:2023A&A...679A.109C}. The relative rate of runaway star production from the binary supernova scenario \citep{Blaauw+1961, Boersma+1961} and the dynamical ejection scenario \citep{Poveda+1967} remains debated \citep[e.g.,][]{DorigoJones+2020:2020ApJ...903...43D, Sana+2022}. The ejection velocity of a runaway star is not strongly dependent on its companion's SN kick, as we will show here. However, the fraction of binaries that are disrupted by SNe is highly sensitive to the distribution of SN kicks. Therefore, SN kicks strongly affect the rate and demographics of runaway stars, and hypervelocity stars \citep[e.g.,][]{Tauris+2015:2015MNRAS.448L...6T}, produced by the binary supernova scenario \citep[e.g.,][]{DeDonder+1997:1997A&A...318..812D,Eldridge+2011:2011MNRAS.414.3501E,Renzo+2019:2019A&A...624A..66R,Wagg+2025:2025arXiv250417903W}.

Binary stellar population synthesis codes are used to make predictions for these binary products that, in tandem with observations, are used to constrain models for SN kicks. These codes apply different choices of SN kick methodologies in their evolution, which have the potential to significantly change the subsequent evolution of binary systems after a SN. In this paper, we compare SN kick prescriptions of three different codes to ensure consistency. In particular, we focus on how the ejection velocities of stellar companions compare for a binary with identical initial conditions evolved in different codes.

\section{Methods}\label{sec:methods-init-conditions}

\subsection{Population synthesis codes and default settings}

We use three different open-source population synthesis codes in our analysis. The \cosmic population synthesis code \citep{COSMIC} applies the method from \citet{pfahl_comprehensive_2002} (though previously followed \citet{Kiel+2009:2009MNRAS.395.2326K}, see Section~\ref{sec:kiel}). Similarly, \compas \citep{COMPAS, Compas+2022:2022JOSS....7.3838C} uses the method of \citet{pfahl_comprehensive_2002}. In contrast, \binc \citep{Izzard+2004:2004MNRAS.350..407I, Izzard+2006:2006A&A...460..565I, Izzard+2009:2009A&A...508.1359I, Izzard+2018:2018MNRAS.473.2984I, Izzard+2023:2023MNRAS.521...35I, Hendriks+2023:2023JOSS....8.4642H} applies \citet{Tauris+1998:1998AA...330.1047T}. The combination of these codes therefore allows us to cross-validate three implementations of two different algorithms for determining secondary ejection velocities.

\begin{table}[tb]
    \centering
    \begin{tabular}{c|l}
        \hline
        \hline
        Code & Supernova kick prescription \\
        \hline
        \multirow{2}{*}{\cosmic} & \citealt{Kiel+2009:2009MNRAS.395.2326K} ($<$v3.5.0) \\
         & \citealt{pfahl_comprehensive_2002} ($\ge$v3.5.0) \\
        \compas & \citealt{pfahl_comprehensive_2002}\\
        \binc & \citealt{Tauris+1998:1998AA...330.1047T}
    \end{tabular}
    \caption{The prescription for supernova natal kicks used in the population synthesis codes we consider.}
    \label{tab:kick-prescriptions}
\end{table}

These codes share a joint origin of BSE \citep{Hurley+2002}, but have since diverged and implemented additional physics, with alternate default settings. As a result, the same binary evolved in each code can result in moderately different evolution. Of particular relevance to this study are the choices regarding remnant mass prescriptions and SN natal kick magnitudes. By default, \cosmic and \binc both use the delayed remnant mass prescription from \citet{Fryer+2012:2012ApJ...749...91F}, which is a deterministic function of the pre-SN CO core mass. In contrast, \compas instead applies the \citet{Mandel+2020} prescription, which is a probabilistic model that accounts for some of the intrinsic stochasticity in stellar evolution and SNe. This model also draws a natal kick based on the remnant mass. \cosmic instead samples a natal kick magnitude from a Maxwellian distribution, with a central parameter of $\sigma = 265\unit{km}{s^{-1}}$ for core-collapse SNe \citep[e.g.,][]{Hobbs+2005:2005MNRAS.360..974H}, and $\sigma = 20\unit{km}{s^{-1}}$ for electron-capture SNe \citep[e.g.,][]{Igoshev+2020}. The default choice for \binc is to sample natal kicks from a single Maxwellian distribution, with a central parameter of $\sigma = 190 \unit{km}{s^{-1}}$ \citep{1997MNRAS.291..569H} and to follow the remnant mass prescription from BSE \citep{Hurley+2000:2000MNRAS.315..543H, Hurley+2002}.

Given that our goal is to compare the implementation of different kick prescription algorithms, we enforce consistent choices for natal kick magnitudes and remnant masses to better isolate any differences between the codes. Therefore, for each code, we draw natal kick magnitudes from a Maxwellian with $\sigma = 265 \unit{km}{s^{-1}}$ and apply the \citet{Fryer+2012:2012ApJ...749...91F} delayed remnant mass prescription. All other settings we leave unchanged from the code's default choices. As a result, the pre-SN evolution in each code is different for the same initial conditions, which we discuss in Section~\ref{sec:pre-sn}.

\subsection{Initial conditions}

We evolve a representative mass-transferring massive binary with a range of random supernova natal kicks in each code. This binary has the following initial conditions: primary mass $m_1 = 20 \msun$, secondary mass $m_2 = 15 \msun$, orbital period $P = 100 \unit{days}$, eccentricity $e = 0$, and metallicity $Z = 0.02$. We repeat the evolution 50,000 times, each with a different random natal kick magnitude and direction. The kick is assumed to be isotropic, and magnitudes are drawn from a Maxwellian with $\sigma = 265 \unit{km}{s^{-1}}$.

This system is representative of O-type stars that undergo case B mass transfer, which often produce ejected secondary stars. Moreover, case B mass transfer is the most common form of mass transfer in this regime \citep[e.g.,][]{Kippenhahn+1967:1967ZA.....65..251K,vandenHeuvel+1969:1969AJ.....74.1095V}, and this system produces an ejected stellar companion with a typical walkaway star velocity \citep[e.g.,][]{Renzo+2019:2019A&A...624A..66R}.

\section{Results}\label{sec:ejection-comparison}

In this Section we report the pre-SN evolution of the binary (Section~\ref{sec:pre-sn}), compare the ejection velocities of secondary stars as calculated by different population synthesis codes (Section~\ref{sec:compare_ejection_codes}), and explain the relationship between natal kicks and companion ejection velocities in detail (Section~\ref{sec:relationship}). 

\subsection{Pre-supernova evolution}\label{sec:pre-sn}

The typical evolution of the system is to initiate stable case B mass transfer and unbind after the primary supernova. This evolution is qualitatively consistent across each code, but the quantitative details vary.

\begin{figure}
    \centering
    \includegraphics[width=\columnwidth]{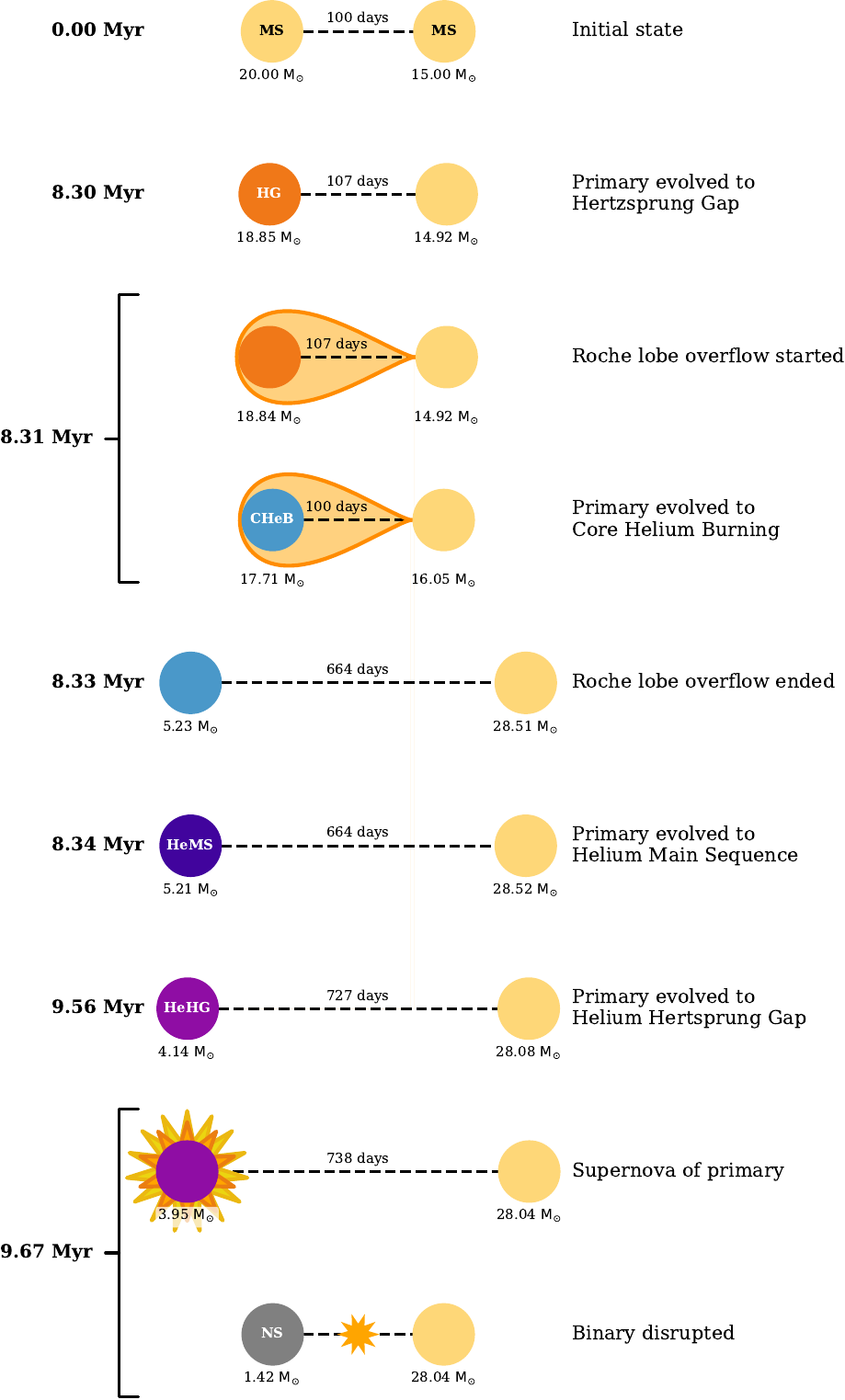}
    \caption{A cartoon evolution diagram of the representative binary that we consider. Each star is represented by a circle, where the colour represents the stellar type and their separation is scaled by their orbital period. Each row indicates an important timestep from a \cosmic simulation, which is labelled with its time and a description to the sides. Acronyms indicate the stellar type, including main sequence (MS), Hertzsprung gap (HG), core helium burning (CHeB), helium main sequence (HeMS), helium Hertzsprung gap (HeHG) and neutron star (NS). Figure dynamically generated with \texttt{cogsworth} \citep{cogsworth_joss, cogsworth}}
    \label{fig:cartoon-evolution}
\end{figure}

We illustrate the typical evolution of this binary in Figure~\ref{fig:cartoon-evolution}, for the system evolved in \cosmic. The primary star initiates almost fully-conservative mass transfer after $8.3\unit{Myr}$, as it expands on the Hertzsprung gap. This transfers $13.6\unit{M_\odot}$ to the secondary star (increasing its mass to $28.5\unit{M_\odot}$) and widens the orbital period from ${\sim}100\unit{d}$ to ${\sim}660\unit{d}$. The primary proceeds to self-strip via stellar winds, further widening the binary until it reaches core collapse $1.5\unit{Myr}$ later.

The difference in binary physics settings between the codes results in moderately different pre-SN evolutionary states. In \compas the evolution proceeds similarly, except the mass transfer is much less conservative, such that a large fraction of the transferred mass is lost from the system. As a result, after the mass transfer ends, the companion is much less massive (${\sim}18\unit{M_\odot}$). Therefore, even though the binary is slightly wider (${\sim}800\unit{d}$) due to the angular-momentum loss from the system, the secondary star's orbital velocity increases. For \binc, the mass transfer is almost identical to \cosmic; however, the stellar winds on both stars are weaker. Consequently, the binary remains tighter, with $P_{\rm orb} = 555\unit{d}$ at the end of mass transfer, and each star has a higher mass ($5.3\unit{M_\odot}$ and $29.3\unit{M_\odot}$ respectively), placing the average ejection velocity between that of \cosmic and \compas. It may be possible to enforce agreement between codes through a detailed selection of binary physics settings but that is beyond the scope of this work.

\subsection{Ejection velocity comparison}\label{sec:compare_ejection_codes}

\begin{figure*}[htb]
    \centering
    \includegraphics[width=0.45\textwidth]{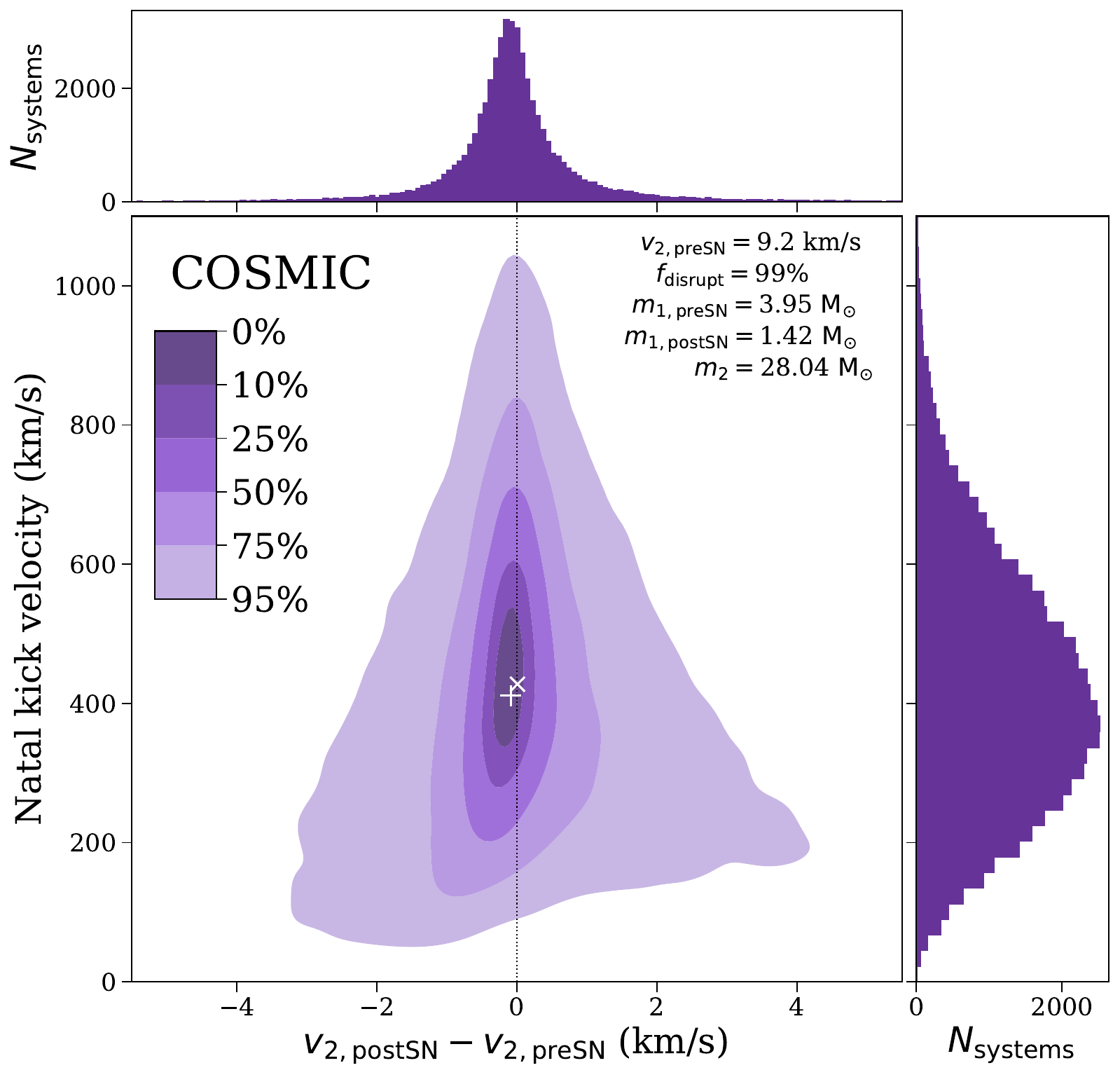}
    \includegraphics[width=0.45\textwidth]{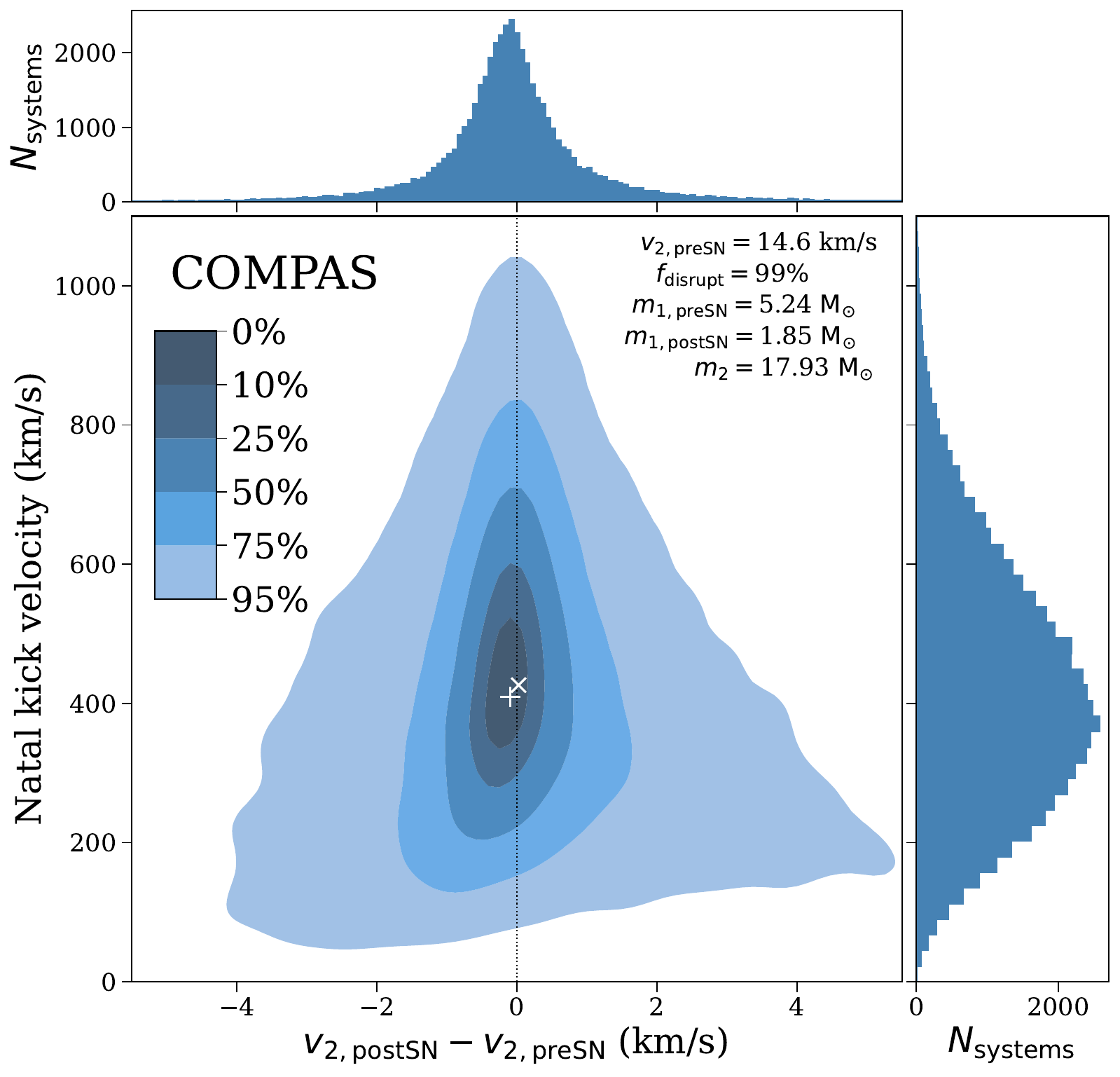}
    \includegraphics[width=0.45\textwidth]{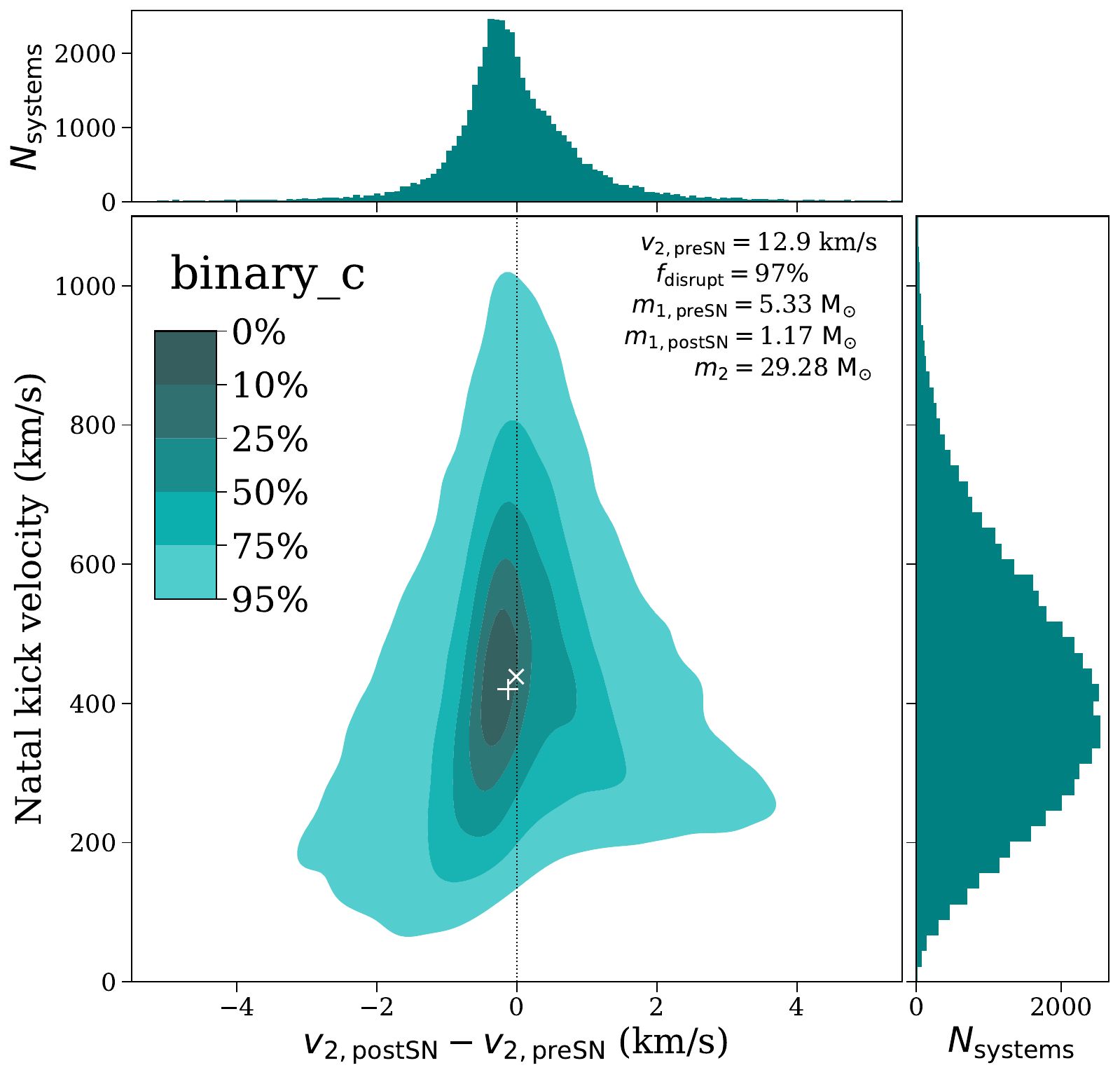}
    \caption{Secondary ejection velocities closely follow pre-SN orbital velocities, with extremely weak dependence on natal kick magnitudes, across different population synthesis codes. These plots compare a binary with identical initial conditions ($m_1 = 20 \, {\rm M_\odot}, \ m_2=15 \, {\rm M_\odot}, \ P_{\rm orb} = 100 \, {\rm days}, \ e = 0, \ Z = 0.02$), evolved 50,000 times with different random natal kicks in three codes. For each code, the main panel shows shows the difference between the secondary ejection velocity and its pre-SN orbital velocity on the $x$-axis, while the $y$-axis shows the magnitude of the natal kick applied to the primary star. Each main panel shows a 2D kernel density estimator with a colourbar indicating the fraction of the population contained within each contour. Secondary panels show marginal histograms. Small white crosses and pluses indicate the mean and median of each distribution respectively. Panels are annotated with the population synthesis code used for evolution and information about the pre- and post-SN system.}
    \label{fig:ejection_vels}
\end{figure*}

The ejection velocities of secondary stars closely follow their pre-SN orbital velocities in each code, and have only a weak dependence on the magnitude of natal kicks. In Figure~\ref{fig:ejection_vels}, we show how the applied natal kick affects the difference between the pre- and post-SN velocity of the secondary star in these simulations. For a circular binary, the pre-SN orbital velocity of the secondary star, $v_{\rm 2, preSN}$, is given by:
\begin{equation}
    v_{\rm 2, preSN} = \frac{m_1}{m_1 + m_2} \cdot v_{\rm orb},
\end{equation}
where $m_1$ is the primary mass, $m_2$ is the secondary mass, and
\begin{equation}
    v_{\rm orb} = \sqrt{\frac{G (m_1 + m_2)}{a}},
\end{equation}
where $G$ is the gravitational constant and $a$ is the semi-major axis. We note that any post-mass-transfer system is typically assumed to be circularised \referee{at periastron separation} in population synthesis, as is the case for all three codes in this study. The main panels show a 2D kernel density estimator for the disrupted binaries in the 50,000 realisations, while the side panels show marginal histograms for each axis.

The distributions across codes are very similar despite their different pre-SN evolution. The mean ejection velocity in each code (indicated by the white crosses) aligns directly with the pre-SN orbital velocity. Quantitatively, we find that, for each code, the ejection velocity of the secondary star is within ${\sim}5\unit{km}{s^{-1}}$ of its pre-SN orbital velocity, despite the natal kick magnitudes spanning 3 orders of magnitude. 

The weak dependence of the secondary's ejection velocity on the natal kick applied to the primary is expected. Due to the strength of the natal kicks, the secondary star experiences a near-instantaneous disappearance of the compact object from its orbit, removing the centripetal acceleration and leading to its ejection at close to its pre-SN orbital velocity (see \citealt{Tauris+1998:1998AA...330.1047T} for a more in-depth discussion).

In addition to the ejection velocity, the fraction of instances of the binary that are disrupted, $f_{\rm disrupt}$, is very similar. We find that $f_{\rm disrupt} = 99\%$, $99\%$ and $97\%$ for \cosmic, \compas, and \binc respectively. For a fixed kick angle, the likelihood of disruption is driven by the ratio of the kick velocity to the pre-SN orbital velocity \citep[e.g.,][]{Kalogera+1996:1996ApJ...471..352K}. Therefore, the lower disruption fraction for \binc is likely due to the binary being tighter at the onset of core collapse (see Section~\ref{sec:pre-sn}).

The overall agreement between the codes validates their implementation of the SN kick routines. Although each code produces a markedly different pre-SN evolution of the system (compare the annotations in each plot in Figure~\ref{fig:ejection_vels}), the relationship between the pre- and post-SN velocities holds. \referee{Furthermore, we have confirmed that, for an identical pre-SN system, kick magnitude, kick direction, and mean anomaly, the resulting ejection velocity is identical when applying the methods of \citet{Tauris+1998:1998AA...330.1047T} and \citet{pfahl_comprehensive_2002}.} \referee{This is expected because both approaches are equivalent, but with differing derivations and coordinate systems. \citet{Tauris+1998:1998AA...330.1047T} uses a coordinate system defined based on the direction of the companion, whilst \citet{pfahl_comprehensive_2002} opts for a system based on the Laplace-Runge-Lenz vector and instead uses a more compact vector notation.}

\subsection{The relationship between natal kick magnitude, angle and companion ejection velocities}\label{sec:relationship}

\begin{figure*}
    \centering
    \includegraphics[width=0.49\textwidth]{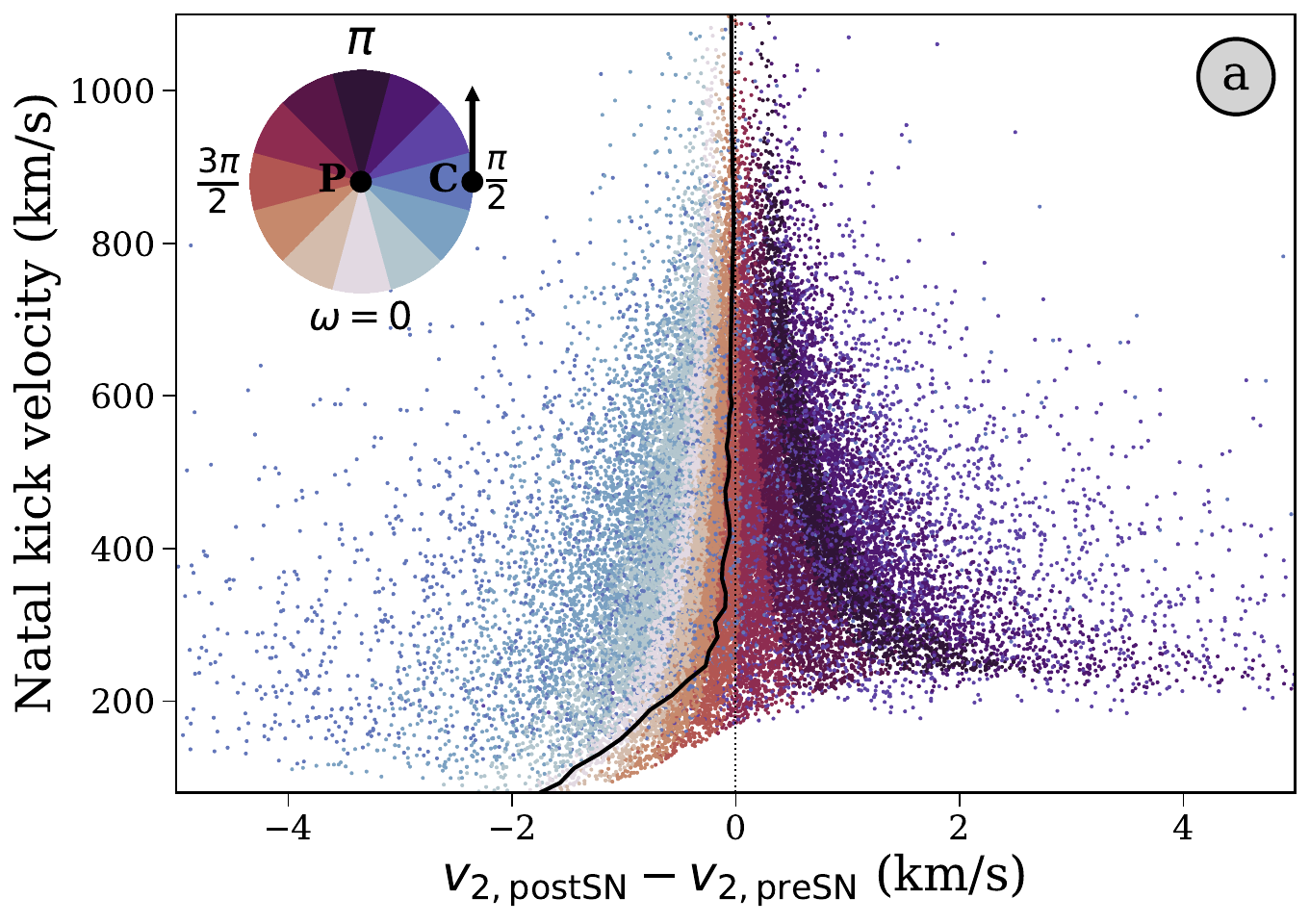}
    \includegraphics[width=0.49\textwidth]{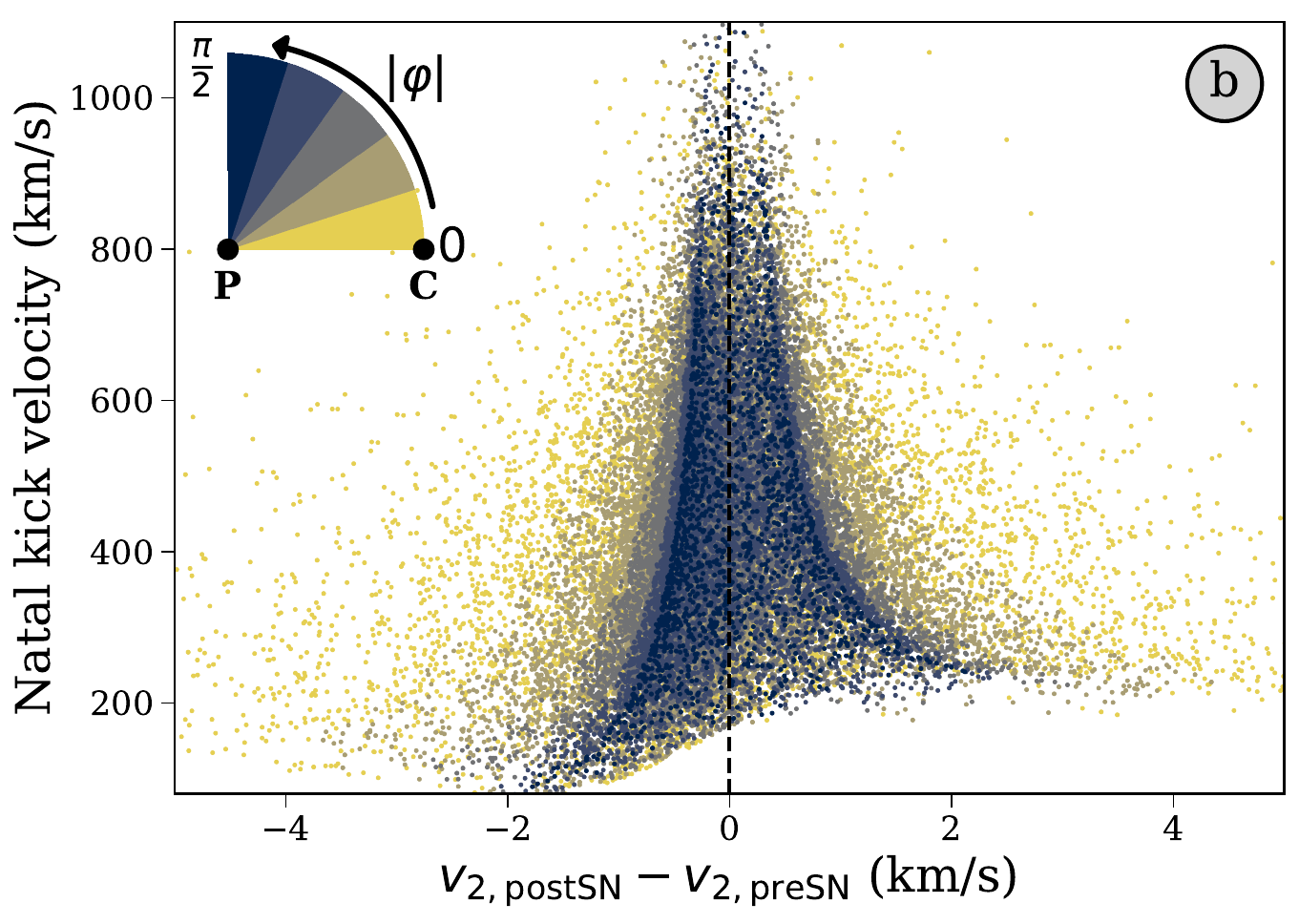}
    \caption{The slight variances in the ejection velocity of the secondary star are due to the angle of the natal kick, which are exaggerated for weaker, in-plane kicks. \textbf{Left (azimuthal angle):} The scatter points show the 50,000 different realisations of the binary run with \binc, each of which are coloured by the azimuthal angle of the kick $\omega$. The solid black line shows the median difference between the pre- and post-SN velocity of the secondary star as a function of the natal kick magnitude. In the upper left corner we include a guide for the geometry of the system, where P is the primary (exploding) star, C is the (ejected) companion star, the arrow indicates the pre-SN orbital velocity of the companion and the wheel of colour indicates the azimuthal angle of the natal kick applied to the primary. The dearth of points in the lower right is where systems remain bound for weak natal kicks with an angle counter to the compact object's orbital motion. \textbf{Right (polar angle):} A similar plot for the absolute value of the polar angle of the kick $\varphi$, where $\varphi = 0$ indicates an in-plane kick. In the schematic in this panel the orbital motion of the stars is directed in and out of the page.}
    \label{fig:angles}
\end{figure*}

The shape of the distributions in Figure~\ref{fig:ejection_vels} can be understood through a consideration of the magnitude and angle of the kick, in concert with the geometry of the system, as shown by several earlier works. In Figure~\ref{fig:angles}, we show the individual disrupted systems from the \binc simulation as scatter points. In the left panel, each point is coloured by the azimuthal angle of the kick $\omega$, while the right panel is coloured by the polar angle of the kick $\varphi$. We include a schematic for the geometry of the system in the upper left corners, which shows the direction of ejection of the companion with an arrow and a coloured wheel indicating the value of $\omega$. \referee{The reference frame is such that $\omega = \pi / 2$ points from the primary star to the companion, and $\varphi = 0$ is within the plane of the binary orbit}. As we noted above, the ejection velocity distribution is centred on the pre-SN orbital velocity of the companion star.

There are several features in Figure~\ref{fig:angles}a that require explanation: (1) the scatter around the mean ejection velocity narrows for strong natal kicks, (2) certain azimuthal angles produce higher or lower ejection velocities, and (3) for weaker kicks, there is a dearth of systems with ejection velocities above the pre-SN orbital velocity and a general skew to lower ejection velocities.

Firstly, stronger natal kicks lead to a tighter relationship between the pre- and post-supernova velocities of the secondary star. This tightening is because a stronger natal kick more immediately removes the compact object from the system. As a result, the compact object is less able to influence the trajectory of the ejected companion star.

Secondly, the azimuthal angle of the kick influences the difference between ejection velocity and the pre-supernova velocity of the secondary star. Figure~\ref{fig:angles}a indicates that the strongest accelerations of the companion occur for angles close to $\omega = 5\pi/6$ (light purple), while the strongest decelerations occur for angles close to $\omega = \pi/3$ (blue). From the geometry of the system, we can see that these angles respectively correspond to a trajectory that leads to the compact object being mostly ahead, or mostly behind, the companion ejection direction. For a compact object that is ahead of the direction of the companion, its will pull the companion towards it, hence the acceleration at these angles, while the inverse reasoning applies for the deceleration. For the same reason, natal kicks pointed directly away from the companion ($\omega = 3\pi / 2$) have the weakest effect on the ejection velocity of the companion.

\pagebreak

Lastly, for weaker kicks ($\lesssim 250 \unit{km}{s^{-1}}$) the shape of the distribution shifts because the approximation that the compact object near-instantaneously leaves the system no longer holds. In this regime, a kick directed away from the companion, or behind its ejection trajectory (i.e.\ $\omega < \pi/2$ or $\omega > 3\pi/2$) can allow the compact object to more significantly decelerate the companion before leaving the system. This is notable in the shift of the distribution to more negative values, meaning an ejection slower than the pre-SN orbital velocity of the companion, for weak kicks. Furthermore, a weaker kick with an angle counter to the compact object's orbital motion (i.e.\ $\pi / 2 \le \omega \le 3\pi/2$) can simply decelerate the compact object, tightening the binary, but leaving it bound. The binary remains bound in this way only for approximately 1\%, 1\% and 3\% of kicks applied to the system, for \cosmic, \compas and \binc respectively. The bound systems leave a gap in the distribution of disrupted systems in the lower right of Figure~\ref{fig:angles}a. As a result of these effects, the median difference between the pre- and post-SN velocity of the secondary star is negative for weak kicks, as shown by the solid black line in Figure~\ref{fig:angles}a.

Each of these effects are most prominent for kicks that are directed within the orbital plane of the binary. In Figure~\ref{fig:angles}b, we instead colour each point by the angle of the kick relative to the orbital plane. In-plane kicks ($\varphi = 0$, yellow) produce the full range of deviations from the pre-SN orbital velocity. These kicks can effect a greater change on the ejection velocity because they allow the compact object to remain closer to the companion, and thereby exert a greater gravitational influence, for a longer period of time. For kicks directed perpendicular to the orbital plane ($|\varphi| = \pi / 2$, dark blue), the post-SN influence of the compact object is always reduced. Therefore, the maximal deviation from the pre-SN orbital velocity is decreased, with almost no perpendicular kick instances deviating by more than $2\unit{km}{s^{-1}}$.

\section{Correction to Kiel \& Hurley (2009)}\label{sec:kiel}

Earlier versions of \cosmic ($<$v3.5.0) used the methodology described by \citet{Kiel+2009:2009MNRAS.395.2326K}, hereafter \citetalias{Kiel+2009:2009MNRAS.395.2326K}, in calculating the post-SN orbital properties of binaries, as well as the ejection velocities of compact objects and their prior companions. However, when using this method, we found that the ejection velocities of companions were not weakly correlated with their pre-SN orbital velocities as one would expect. Instead, their ejection velocity showed a strong  (unphysical) dependence on the natal kick received by the primary. Moreover, the ejection velocity of the compact object was often significantly slower than the natal kick that it received.

\begin{figure}
    \centering
    \includegraphics[width=\columnwidth]{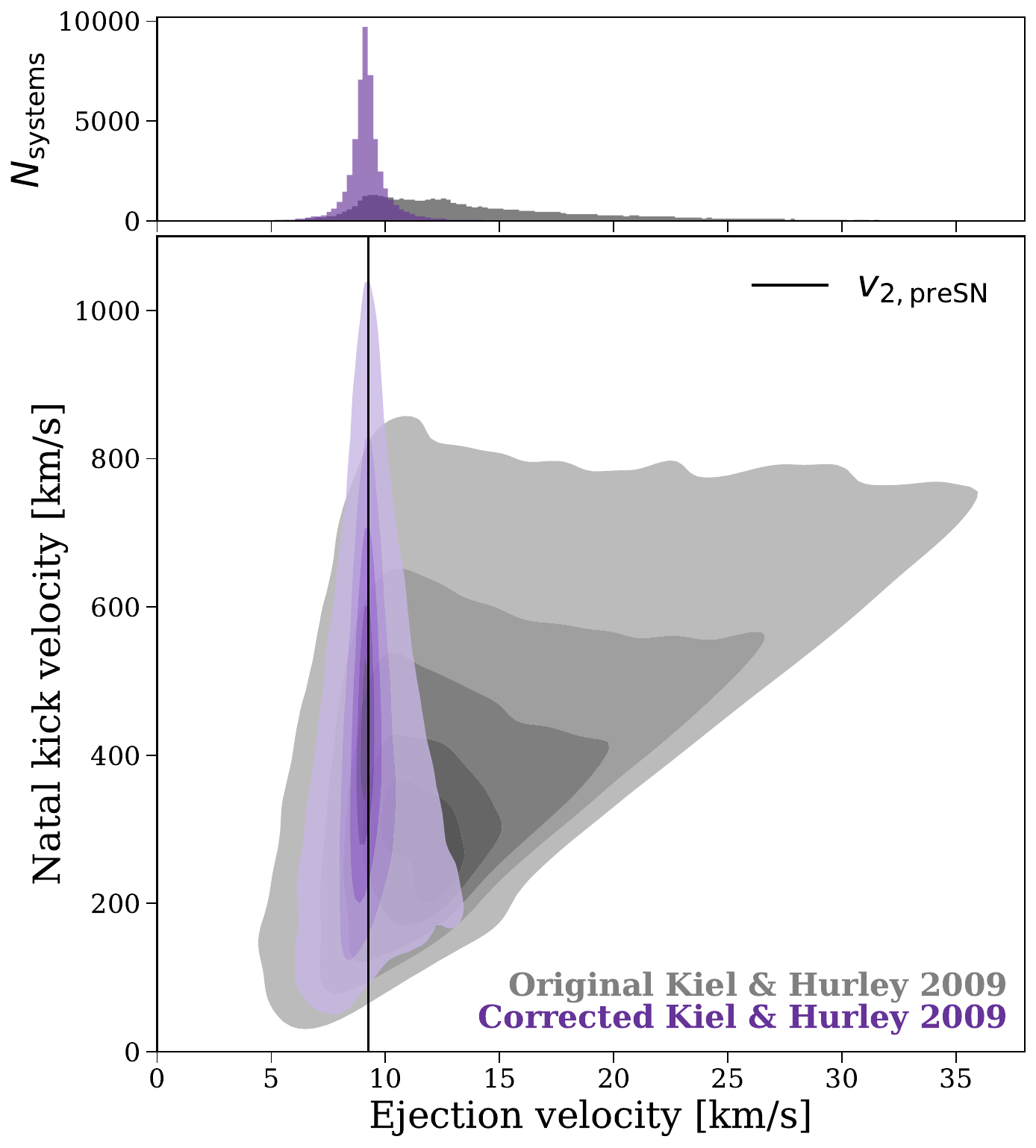}
    \caption{Our correction to \citet{Kiel+2009:2009MNRAS.395.2326K} removes an unphysical dependence of the ejection velocity of secondary stars on the magnitude of the primary star's natal kick. The main panel shows kernel density estimators for the 2D density distributions run in \cosmic, with and without our correction. The upper panel shows the marginal distribution for the ejection velocity.}
    \label{fig:kiel_comparison}
\end{figure}

We demonstrate the (unphysical) strong dependence of the ejection velocity on natal kick magnitude in Figure~\ref{fig:kiel_comparison}. The grey contour plot shows the distribution of ejection velocities and natal kick magnitudes for our representative binary, evolved 50,000 times using the original \citetalias{Kiel+2009:2009MNRAS.395.2326K} prescription. The triangular shape of the contour indicates that stronger natal kicks also lead to faster ejection velocities. Moreover, the marginal distribution in the upper panel of Figure~\ref{fig:kiel_comparison} shows there is a wide spread in ejection velocities, extending up to 4 times the pre-supernova orbital velocity. This dependence is unphysical, as the ejection velocity of the companion should be closely related to its pre-SN orbital velocity.

The reason for these effects is that, in \citetalias{Kiel+2009:2009MNRAS.395.2326K}, the $z$-component of the final velocity of each binary constituent neglects the contributions from the velocity at infinity of the hyperbolic orbit, $V_\infty$. The derivation assumes that this velocity is directed entirely within the orbital plane of the binary, but does not fully account for the fact that the orbital plane is no longer aligned with the $z$ axis after the SN. Therefore, the $z$-component of the final velocity of the compact object (equation 17 from \citetalias{Kiel+2009:2009MNRAS.395.2326K}) should be as follows

\begin{equation}
    V_{\rm 1fz} = \frac{M_{\rm NS}}{M_b^{\prime}} V_{\rm kick} \sin \phi \ \underbrace{ + V_{\infty} \sin \nu \sin \gamma \frac{M_2}{M_b^\prime}}_{\rm Term \, missing \, from \, KH09},
\end{equation}
where the variables are defined in the same way as \citetalias{Kiel+2009:2009MNRAS.395.2326K}, such that $M_{\rm NS}$ is the mass of the resulting compact object, $M_b^\prime$ is the final total mass of the system, $V_{\rm kick}$ is the magnitude of natal kick, $\phi$ is the polar angle of the natal kick, $V_\infty$ is the velocity at infinity of the hyperbolic orbit and $\nu$ and $\gamma$ are angles accounting for the misalignment of the orbital plane after the SN (see Section 2.2.1 and Figure 2 of \citetalias{Kiel+2009:2009MNRAS.395.2326K}). Similarly, the $z$-component of the final velocity of the companion (equation 20 from \citetalias{Kiel+2009:2009MNRAS.395.2326K}) should be as follows
\begin{equation}
    V_{\rm 2fz} = \frac{M_{\rm NS}}{M_b^{\prime}} V_{\rm kick} \sin \phi \ \underbrace{- V_{\infty} \sin \nu \sin \gamma \frac{M_{\rm NS}}{M_b^\prime}}_{\rm Term \, missing \, from \, KH09}.
\end{equation}

We have confirmed that including the additional $V_\infty$ terms in these equations removes the unphysical dependence on natal kick magnitude. In Figure~\ref{fig:kiel_comparison}, the purple distributions correspond to the corrected version of \citetalias{Kiel+2009:2009MNRAS.395.2326K}, which now strongly correlates with the pre-supernova orbital velocity. This change brings the \citetalias{Kiel+2009:2009MNRAS.395.2326K} method into agreement with that of \citet{pfahl_comprehensive_2002} and \citet{Tauris+1998:1998AA...330.1047T}.

The latest versions of \cosmic ($\ge$v3.5.0) now use the \citet{pfahl_comprehensive_2002} method by default. We make the original \citet{Kiel+2009:2009MNRAS.395.2326K} method accessible (via negative \texttt{kickflag} inputs) so that earlier \cosmic results can be reproduced. We note that this issue only affected the ejection velocities of stars and compact objects from disrupted systems, but the \textit{conditions} for disruption and post-SN orbital parameters for \textit{bound} systems were correct. Therefore prior work using \cosmic to explore bound systems remains unchanged.

\section{Summary and importance of open-science comparisons}

We have compared the ejection velocities of stellar companions for a representative case B mass-transferring binary across three independent open-source rapid population synthesis codes. We have confirmed that each code is consistent in its finding that there is only a very weak dependence of the stellar ejection velocity on the SN natal kick, as one would physically expect. We explain the influence of the direction and magnitude of natal kicks in slightly shifting the companion's ejection velocity from its pre-SN orbital velocity and provide a correction to an earlier prescription for SN kicks.

It is notable that, during the course of this project, we identified problems with either the kick prescription or implementation in each of the codes that we considered. As discussed above, \cosmic previously used an erroneous kick prescription, which has been resolved as of v3.5.0\footnote{\url{https://github.com/COSMIC-PopSynth/COSMIC/pull/678}}. The implementation in \compas had anomalous component velocities such that the ejection velocity of secondary stars was consistently overestimated by a factor of 2. This problem arose as a result of spurious vector indexing and incorrect rotation algebra in the code and has been resolved as of \compas v03.09.02\footnote{\url{https://github.com/TeamCOMPAS/COMPAS/pull/1290}}. In \binc, the prior implementation did not sample natal kick magnitudes and azimuthal angles in an independent manner, which led to slightly lower ejection velocities on average. This correlated sampling occurred as a result of re-using a randomly generated number from the kick magnitude in the azimuthal angle sampling. This issue has been resolved in \binc for versions after v2.24\footnote{\url{https://gitlab.com/binary_c/binary_c/-/commit/fba20687984e9cfb46bc38838d07655bfc99c6e5}}.

Despite the rigorous testing suites of each code that aim to address issues such as these, each bug was not uncovered until comparison with other open-source codes. It is therefore evident that open science, and comparison projects \citep[e.g.,][]{Toonen+2014:2014A&A...562A..14T, O'Connor+2018:2018JPhG...45j4001O}, are essential to identifying and resolving subtle issues in population synthesis.\\\\\\\\


\section*{Data Availability}
All data and code to reproduce these results are available \href{https://doi.org/10.5281/zenodo.15225730}{on Zenodo}\footnote{\url{https://doi.org/10.5281/zenodo.15225730}} and in a \href{https://github.com/TomWagg/ejection-velocities/}{GitHub repository}\footnote{\url{https://github.com/TomWagg/ejection-velocities/}}.

\section*{Acknowledgements}
\referee{We thank the referee, Jeff Andrews, for his helpful comments that improved this work. We additionally thank Johnny Dorigo Jones, Adam Burrows, and Thomas Tauris for their valuable comments on this work.} We are grateful to Eric Bellm, Andy Tzanidakis, David Wang, and John Franklin Crenshaw for helpful discussions of supernova kick coordinate systems. TW and KB acknowledge support from NASA ATP grant 80NSSC24K0768.

\textit{Software:} This work made use of the following software packages: \texttt{astropy} \citep{astropy:2013, astropy:2018, astropy:2022}, \texttt{Jupyter} \citep{2007CSE.....9c..21P, kluyver2016jupyter}, \texttt{matplotlib} \citep{Hunter:2007}, \texttt{numpy} \citep{numpy}, \texttt{pandas} \citep{mckinney-proc-scipy-2010, pandas_13819579}, \texttt{python} \citep{python}, \texttt{scipy} \citep{2020SciPy-NMeth, scipy_13352243}, \texttt{COSMIC} \citep{Breivik2020, COSMIC_13854578}, \binc \citep{Izzard+2004:2004MNRAS.350..407I, Izzard+2006:2006A&A...460..565I, Izzard+2009:2009A&A...508.1359I, Izzard+2018:2018MNRAS.473.2984I, Izzard+2023:2023MNRAS.521...35I, Hendriks+2023:2023JOSS....8.4642H}, \texttt{Cython} \citep{cython:2011}, \texttt{h5py} \citep{collette_python_hdf5_2014, h5py_7560547}, \texttt{schwimmbad} \citep{schwimmbad}, \texttt{seaborn} \citep{Waskom2021}, and \texttt{tqdm} \citep{tqdm_13207611}. This research has made use of NASA's Astrophysics Data System. Simulations in this paper made use of the COMPAS rapid binary population synthesis code (version 3.01.10), which is freely available at \url{http://github.com/TeamCOMPAS/COMPAS} \citep{COMPAS}. Software citation information aggregated using \texttt{\href{https://www.tomwagg.com/software-citation-station/}{The Software Citation Station}} \citep{software-citation-station-paper, software-citation-station-zenodo}.


\bibliographystyle{aasjournal}
\bibliography{paper}{}

\end{document}